\documentclass[fleqn,twoside]{article}
\usepackage{espcrc2}

\usepackage{graphicx}

\newcommand{\rr}{\mbox{\boldmath $r$}}

\newcommand{\rb}{\mbox{\boldmath $b$}}

\newcommand{\AmS}{{\protect\the\textfont2
  A\kern-.1667em\lower.5ex\hbox{M}\kern-.125emS}}

\title{Heavy quark photoproduction
       in $pp$ coherent interactions at LHC }

\author{V.P. Gon\c{c}alves\address[MCSD]{Universidade Federal de Pelotas, \\
        P.O. Box 354, Campus Universit\'{a}rio, sn, CEP 96010-900, Pelotas - RS, Brazil},
        M. V. T. Machado\address{ Universidade Federal do Pampa,\\
        P. O. BOX 07, Rua Carlos Barbosa, sn, CEP 96412-420, Bag\'{e} - RS, Brazil\\}
        and
        A.R. Meneses\addressmark[MCSD].}

\begin{document}

\begin{abstract}
In this work we analyse the possibility of constraining  the QCD
dynamics at high energies studying the heavy quark photoproduction
at LHC in coherent interactions. The rapidity distribution and total
cross section for charm and bottom production are estimated using
three different phenomenological saturation models which
successfully  describe the HERA data. Our
results indicate that the experimental study of the inclusive heavy
quark photoproduction can be very useful to discriminate between the
classical and quantum versions of the Color Glass Condensate (CGC) formalism\vspace{1pc}.
\end{abstract}

\maketitle

\section{INTRODUCTION}

The cross sections for heavy quark production in hadron-hadron and
lepton-hadron collisions at high energies are strongly dependent on
the behavior of the gluon distribution, which is determined by the
underlying QCD dynamics. Theoretically, at high energies the QCD
evolution leads to a system with high gluon density, characterized
by the limitation on the maximum phase-space parton density that can
be reached in the hadron wave function (parton saturation). The
transition is specified by a typical scale, which is energy
dependent and is called the saturation scale. Signals of parton
saturation have already been observed both in ep deep inelastic
scattering at HERA and in deuteron-gold  collisions at RHIC.
However, the observation of this new regime still  needs
confirmation and so there is an active search for new experimental
signatures. In this contribution we study the inclusive and
diffractive photoproduction of heavy quarks in coherent
proton-proton collisions considering three phenomenological models
based on the Color Glass Condensate formalism, which describe quite well the
current experimental HERA data for inclusive and exclusive
observables (For a detailed discussion see Ref. \cite{1}). 

In
hadron-hadron colliders, the relativistic protons give rise to
strong electromagnetic fields, which can interact with each other.
Namely, quasi-real photons scatters off protons at very high
energies in the current hadron colliders \cite{2}. Our main motivation
comes from the fact that in coherent interactions at Tevatron and
LHC the photon reaches energies higher than those currently
accessible at DESY - HERA. 
The heavy quark photoproduction cross
section in a proton-proton collision is given by,
\begin{eqnarray}
   \sigma(p+p \rightarrow p+ Q\overline{Q} + Y) = 2 \int_{0}^{\infty}
   \frac{dN_{\gamma}(\omega)}{d\omega} \nonumber \\
  \times  \, \sigma_{\gamma p\rightarrow Q\overline{Q} Y}\left(W_{\gamma p}^2= 2\,\omega\,\sqrt{S_{NN}}  \right) \, d \omega \; ,
\label{eq:sigma_pp}
\end{eqnarray}
where $\omega$ is the photon energy in the center-of-mass frame
(c.m.s.), $W_{\gamma p}$ is the c.m.s. photon-proton energy and
$\sqrt{S_{NN}}$ denotes the proton-proton c.m.s. energy.  The final
state $Y$ can be  a hadronic state generated by the fragmentation of
one of the colliding protons (inclusive production) or a  proton
(diffractive production).

\section{PHOTOPRODUCTION OF HEAVY QUARKS}

In the color dipole approach the inclusive and diffractive heavy
quark photoproduction cross section  are given by \cite{1,3,4}
\begin{eqnarray}
\sigma_{tot}\,   =  2\, \int d^2\rb \,\, \langle \mathcal{N}(x,\rr,\rb) \rangle
\label{totalcsinc}
\end{eqnarray}
and
\begin{eqnarray}
\sigma_{tot}^D\,  = \int d^2\rb \,\, \langle \mathcal{N}^2(x,\rr,\rb) \rangle
\,\,. 
\label{totalcsdif}
\end{eqnarray}
where 
\begin{eqnarray}
 \langle ( ... )  \rangle \equiv \int d^2\rr \int dz
|\Psi_{\gamma}(\rr,z)|^2 \, (...)
\end{eqnarray}
In these equations the variable $\rr$ defines the relative transverse
separation of the pair (dipole), $z\, (1-z)$ is the longitudinal
momentum fractions of the quark  (antiquark) and the function $\Psi_{\gamma}$ is the
light-cone wave function for transversely  polarized photons, which
depends in our case of the charge and mass of the heavy quark. The
function  $N (x,\rr,\rb)$ is the forward dipole-target scattering
amplitude for a dipole with size  $\rr$ and impact parameter $\rb$ which
encodes all the information about the hadronic scattering, and thus
about the non-linear and quantum effects in the hadron wave function
\cite{5}. During the last years an intense activity in the area resulted
in sophisticated models for the dipole-proton scattering amplitude,
which have strong theoretical constraints and which are able to
describe the HERA and/or RHIC data. In what follows we will use
three distinct phenomenological saturation models based on the Color
Glass Condensate formalism which describe quite well the more recent HERA
data: the IIM \cite{6}, the bCGC \cite{7} and the IP-SAT \cite{8} models. In the
bCGC model, which is the quantum version of the CGC formalism,
the dipole-proton scattering amplitude is given by
\begin{eqnarray}
\mathcal{N}^{\mbox{bCGC}} = \left\{ \begin{array}{ll} {\mathcal
N}_0\, \left(\frac{ r \, Q_{s}}{2}\right)^{2\left(\gamma_s +
\frac{\ln (2/r Q_{s})}{\kappa \,\lambda \,Y}\right)}  & \mbox{$r Q_{s} \le 2$} \\
 1 - \exp[{-A\,\ln^2\,(B \, r \, Q_{s})}]   & \mbox{$r Q_{s}  > 2$}
\end{array} \right. \nonumber
\label{eq:bcgc}
\end{eqnarray}
with
\begin{eqnarray}
  Q_{s}=\left(\frac{x_0}{x}\right)^{\frac{\lambda}{2}}\;
\left[\exp\left(-\frac{{\rb}^2}{2B_{\rm
CGC}}\right)\right]^{\frac{1}{2\gamma_s}}.  \label{newqs}
\end{eqnarray}
Furthermore, we will use in our calculations the scattering
amplitude scattering proposed in Ref. \cite{8}, denoted IP-SAT,
which is given by
\begin{eqnarray}
\mathcal{N}^{\mbox{IP-SAT}}(x,\rr,\rb) = 1 - \mathcal{S}(x,\rr,\rb)
\,\,, \label{nkmw}
\end{eqnarray}
where
\begin{eqnarray}
\mathcal{S} =  \exp\left( - \frac{\pi^2}{2\,N_c} \rr^2
\alpha_s(\mu^2)\,xg(x,\mu^2)\,T(\rb)\right),
\end{eqnarray}
 the scale
$\mu^2$ is related to the dipole size $\rr$ by $\mu^2 = 4/\rr^2 +
\mu_0^2$ and the gluon density is evolved from a scale $\mu_0^2$ up
to $\mu^2$ using LO DGLAP evolution without quarks assuming that the
initial gluon density is given by $xg(x,\mu_0^2) = A_g \,
x^{-\lambda_g}\,(1 - x)^{5.6}$. The values of the parameters
$\mu_0^2$, $A_g$ and $\lambda_g$ are determined from a fit to $F_2$
data. Moreover, it is assumed that the proton shape function
$T(\rb)$  has a Gaussian form, $T(\rb) = 1/(2\pi
B_G)\exp[-(\rb^2/2B_G)]$, with $B_G$ being a free parameter which is
fixed by the fit to the differential cross sections for exclusive
vector meson production. The parameter set used in our calculations
is the one presented in the first line of Table III of \cite{8}:
$\mu_0^2 = 1.17$ GeV$^2$, $A_g = 2.55$,  $\lambda_g = 0.020$ and
$B_G = 4$ GeV$^{-2}$. The previous expression  for the forward
scattering amplitude can be obtained to leading logarithmic accuracy
in the classical effective theory of the Color Class Condensate
formalism. Moreover, it is applicable when the leading logarithms in
$Q^2$ dominate the leading logarithms in $1/x$, with the small $\rr$
limit being described by the linear DGLAP evolution at small-$x$. In
contrast, the bCGC model for $\mathcal{N}$
captures the basic properties of the quantum evolution in the CGC
formalism, describing both the bremsstrahlung limit of linear
small-$x$ evolution (BFKL equation) as well nonlinear
renormalization group at high parton densities (very small-$x$).
Consequently, the IP-SAT model can be considered a phenomenological
model for the classical version of the CGC, while the bCGC for the
quantum limit.It is important to emphasize that both models provide
excellent fits to a wide range of HERA data for $x\le 0.01$.
Therefore, the study of  observables which are strongly dependent on
$\mathcal{N}$ is very important to constrain the underlying QCD
dynamics at high energies. In what follows we consider these two
models as input in our calculations of the inclusive and diffractive
heavy quark photoproduction in coherent $pp$ interactions at  LHC energies.

\begin{figure}[t]
\begin{tabular}{cc}
\includegraphics[scale=0.27] {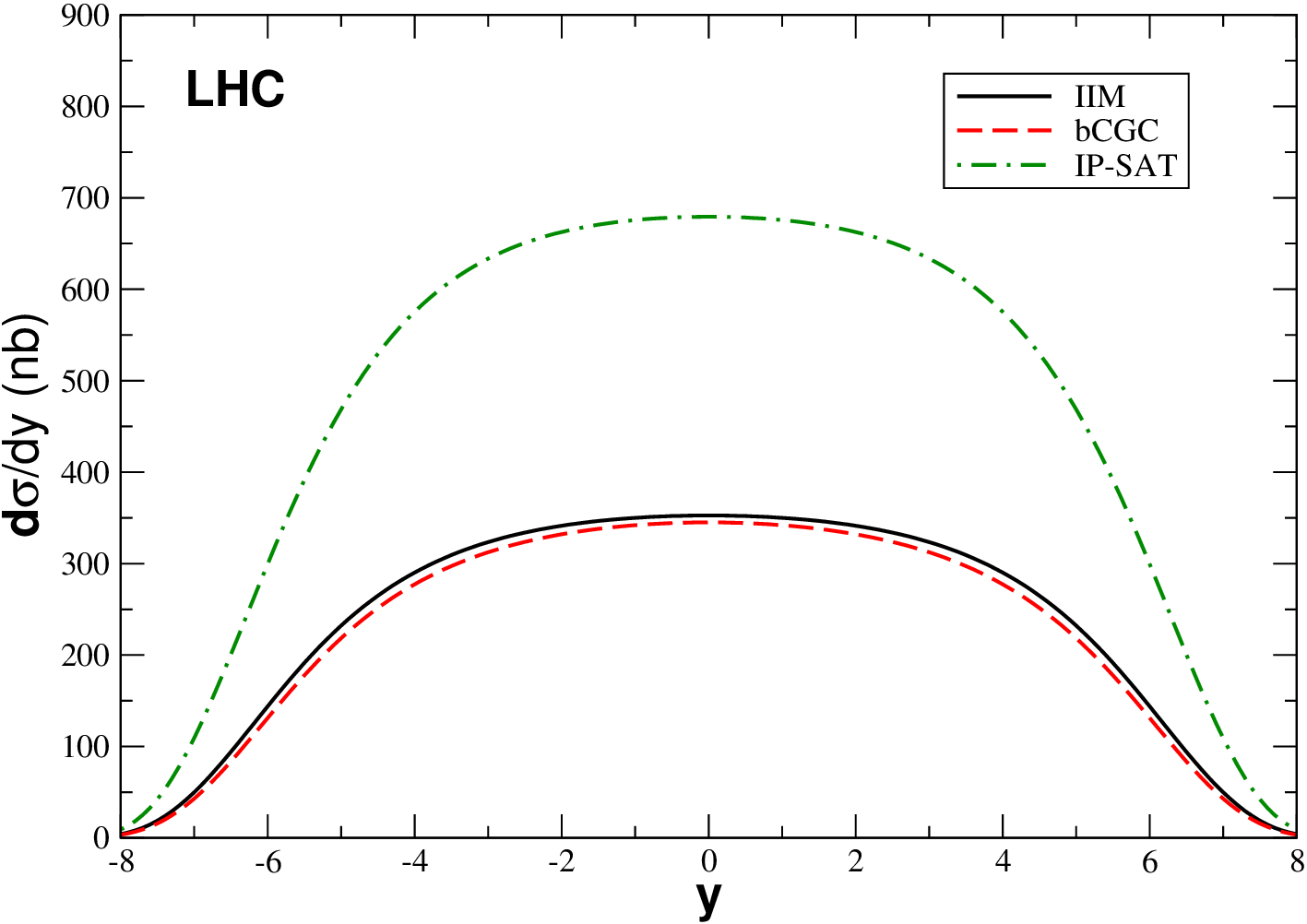} \vspace{1cm} \\  \includegraphics[scale=0.27]{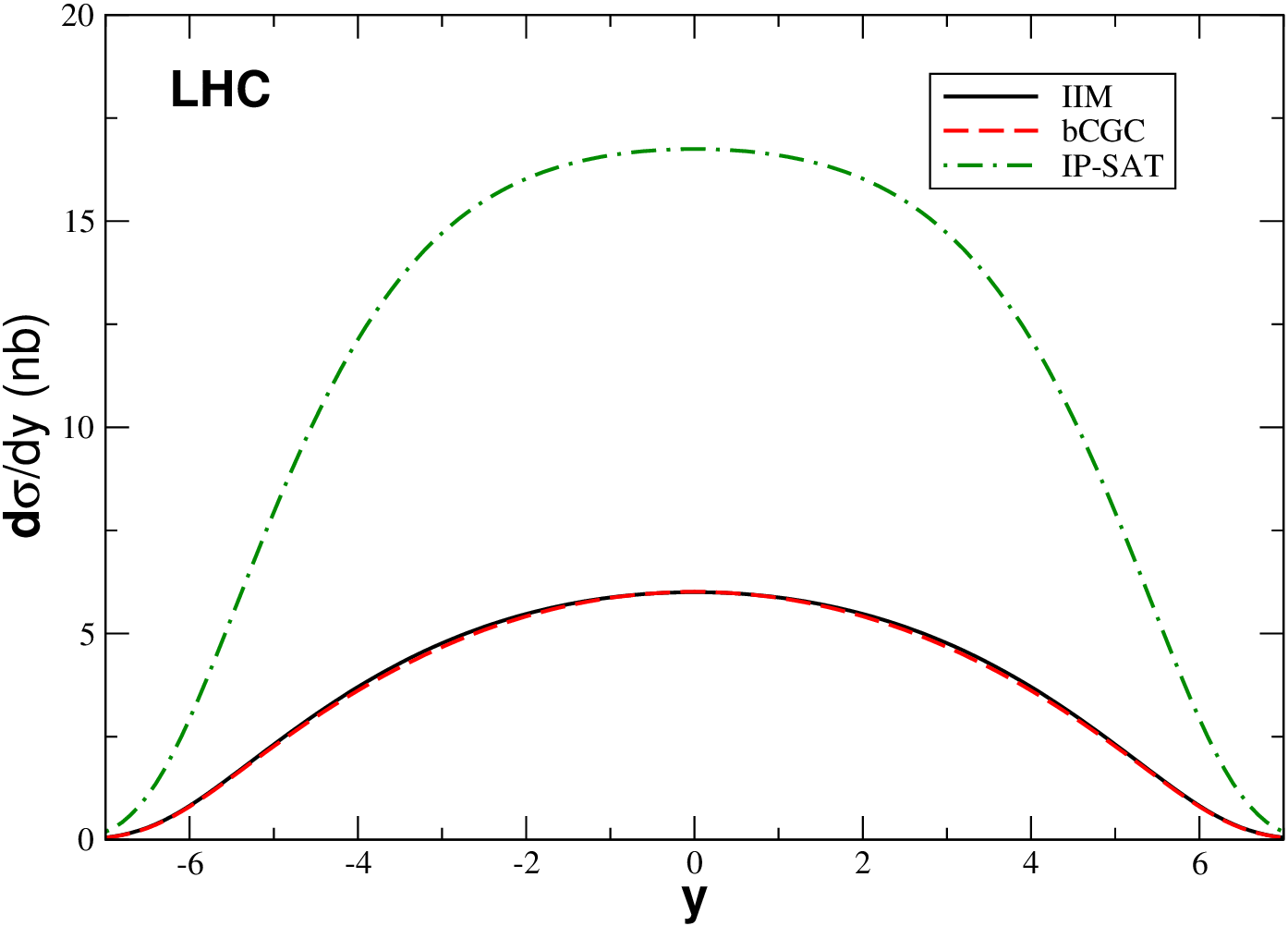}
\end{tabular}
\caption{ The rapidity distribution for the inclusive  charm (upper
panel) and bottom (lower panel) photoproduction on $pp$
reactions at LHC energy $\sqrt{S_{NN}}=14\,\mathrm{TeV}$. Different
curves correspond to distinct phenomenological models.} \label{fig1}
\end{figure}

\section{RESULTS}

The distribution on rapidity $y$ of the produced open heavy quark
state can be directly computed using its relation with the photon
energy $\omega$, i.e, $y\propto ln {\frac{\omega
}{m_Q}}$. A reflection around $y=0$ takes into account the
interchange between the proton emitting the photon  and the target proton. Explicitly, the rapidity distribution is written down as,
\begin{eqnarray}
\frac{d\sigma \,\left[pp \rightarrow pQ\overline{Q}Y)
\right]}{dy} =
 \omega \, \frac{dN_{\gamma} (\omega )}{d\omega
}\,\sigma_{\gamma p \rightarrow Q\overline{Q}Y}, 
\end{eqnarray}
where $Y$ is a hadronic final state $X$ produced by proton
fragmentation in the inclusive case and $Y = p$ for  diffractive
production.

\begin{figure}[t]
\begin{tabular}{cc}
\includegraphics[scale=0.27] {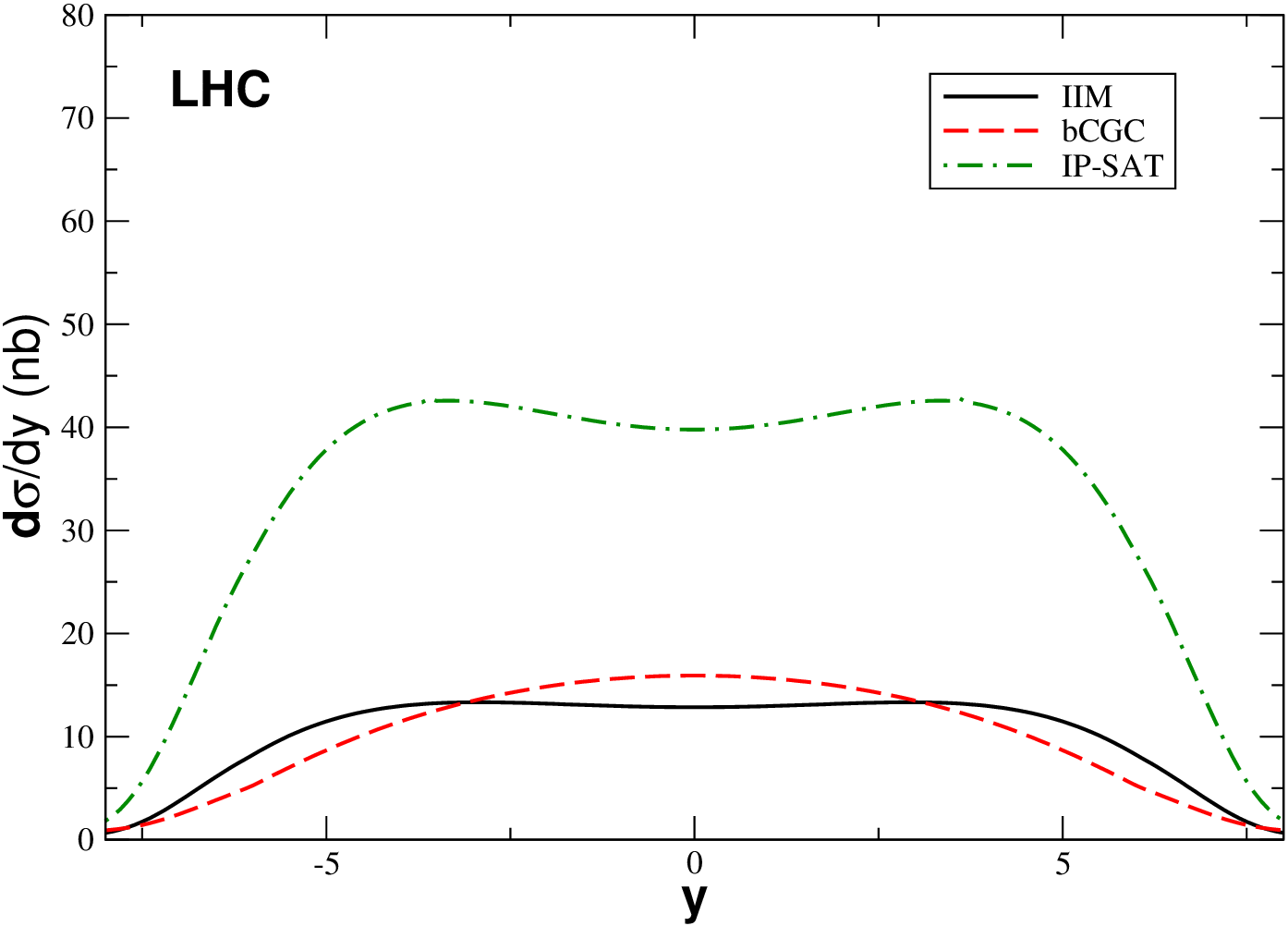} \vspace{1cm} \\  \includegraphics[scale=0.27] {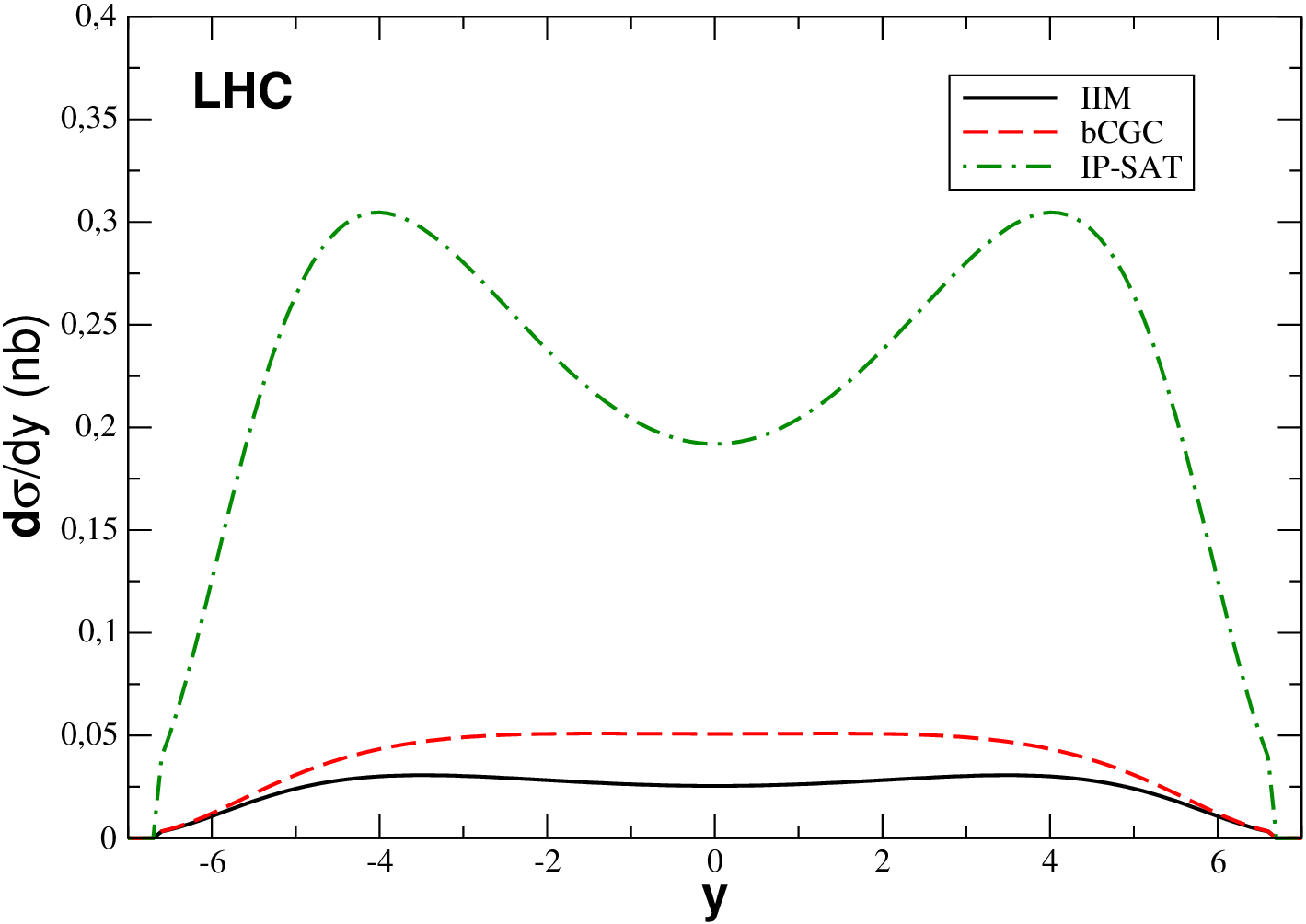}
\end{tabular}
\caption{ The rapidity distribution for the diffractive  charm
(upper panel) and bottom (lower panel) photoproduction on $pp$
reactions at LHC energy $\sqrt{S_{NN}}=14\,\mathrm{TeV}$. Different
curves correspond to distinct phenomenological models.} \label{fig2}
\end{figure}

The resulting rapidity distributions for inclusive and diffractive
heavy quark photoproduction at LHC energies coming out of the
distinct phenomenological models considered in previous section are
depicted in Figs. \ref{fig1} and \ref{fig2}, respectively. For the inclusive case
(Fig. \ref{fig1}), the IIM and bCGC predictions are very similar. In
contrast, these predictions are distinct in the diffractive case
(Fig. \ref{fig2}), with the bCGC prediction being larger than IIM
one at mid-rapidity. On the other hand, the IP-SAT prediction is
larger than these predictions by a factor 2 (3) in the charm
(bottom) case. We can consider the IIM and bCGC predictions as 
lower bounds for the coherent production of heavy quarks at LHC. Our
results indicate that the experimental study of the inclusive heavy
quark photoproduction can be very useful to discriminate between the
classical and quantum versions of the CGC formalism.
It also is true in the diffractive case, where the different models
can be discriminated more easily.

 The results for the integrated cross
section considering the distinct phenomenological models are
presented in Table \ref{tabhq}, for the inclusive and diffractive
charm and bottom pair production at  LHC.
The IP-SAT model gives the  largest rates among the approaches
studied, followed by the bCGC and IIM models with almost identical
predictions. Therefore, these reactions can have high rates at the  LHC kinematical regime.
The cross sections for diffractive production
are approximately two orders of magnitude smaller than the inclusive
case, but due the clear experimental signature of this process (two
rapidity gaps), its experimental analysis still is feasible.

\begin{table}[t]
\begin{center}
\begin{tabular} {||c|c|c|c||}
\hline \hline
 $Q\overline{Q}$   & {\bf IIM} & {\bf bCGC} & {\bf IP-SAT}  \\
\hline \hline
 $c\bar{c} \,\, (\mbox{incl.})$ &  3821 nb & 3662 nb & 7542 nb  \\
\hline
 $c\bar{c} \,\, (\mbox{diff.})$ &  165 nb & 161 nb & 532 nb  \\
\hline
 $b\bar{b} \,\, (\mbox{incl.})$ &  51 nb & 51 nb & 158 nb  \\
\hline
 $b\bar{b} \,\, (\mbox{diff.})$ &  0.32 nb & 0.52 nb & 3 nb  \\
\hline \hline
\end{tabular}
\end{center} \caption{\ The integrated cross section for the
inclusive and diffractive photoproduction of heavy quarks in
$pp$ collisions at  LHC energies.}
\label{tabhq}
\end{table}

\section{CONCLUSIONS}
In summary, we have computed the cross sections for inclusive and
diffractive  photoproduction of heavy quarks in $pp$
collisions at LHC energies. This has been
performed using  modern phenomenological models based on the Color
Glass Condensate formalism, which describe quite well the inclusive
and exclusive observables measured in $ep$ collisions at HERA. The
obtained values are shown to be sizeable  at Tevatron and are
increasingly larger at LHC. The feasibility of detection of these
reactions is encouraging, since their experimental signature should
be sufficiently clear.  Furthermore, they enable to constrain the
underlying QCD dynamics at high energies, which is fundamental to
predict the observables which will be measured in central
hadron-hadron collisions at LHC \cite{9}.

\end{document}